\documentclass{aastex}
\usepackage{emulateapj5}
\usepackage{subfigure}

\begin{document}

\title{Bar-driven dark halo evolution:\\
  a resolution of the cusp--core controversy}
\author{Martin D. Weinberg \& Neal Katz}
\affil{Department of Astronomy, University of Massachusetts, Amherst, MA
  01003-9305}
\email{weinberg@astro.umass.edu, nsk@kaka.astro.umass.edu}

\begin{abstract}
  Simulations predict that the dark matter halos of galaxies should
  have central cusps, while those inferred from observed galaxies do
  not have cusps.  We demonstrate, using both linear perturbation
  theory and n-body simulations, that a disk bar, which should be
  ubiquitous in forming galaxies, can produce cores in cuspy CDM dark
  matter profiles within five bar orbital times.  Simulations of
  forming galaxies suggest that one of Milky Way size could have a 10
  kpc primordial bar; this bar will remove the cusp out to $\sim$5 kpc
  in $\sim$1.5 gigayears, while the disk only loses $\sim$8\% of its
  original angular momentum.  An inner Lindblad-like resonance couples
  the rotating bar to orbits at all radii through the cusp,
  transferring the bar pattern angular momentum to the dark matter
  cusp, rapidly flattening it.  This resonance disappears for profiles
  with cores and is responsible for a qualitative difference in bar
  driven halo evolution with and without a cusp.  This bar induced
  evolution will have a profound effect on the structure and evolution
  of almost all galaxies.  Hence, both to understand galaxy formation
  and evolution and to make predictions from theory it is necessary to
  resolve these dynamical processes.  Unfortunately, correctly
  resolving these important dynamical processes in {\it ab initio}
  calculations of galaxy formation is a daunting task, requiring at
  least 4,000,000 halo particles using our SCF code, and probably
  requiring many times more particles when using noisier tree, direct
  summation, or grid based techniques, the usual methods employed in
  such calculations.

\end{abstract}

\keywords{galaxies:evolution --- galaxies: halos --- galaxies:
  kinematics and dynamics --- cosmology: theory --- dark matter}

\section{Introduction}
\label{sec:intro}

The cold dark matter (CDM) model for structure formation has had a
wide range of successes in explaining the observed universe.  However,
particularly at small scales, several vexing problems remain.  Perhaps
the most widely discussed shortcoming concerns the central density
profiles of galaxies.  Cold dark matter structure formation
simulations predict a universal cuspy halo profile \citep[][hereafter
NFW]{Navarro.Frenk.White:97}.  This profile, $\rho\propto
r^{-\gamma}(1+r/r_s)^{\gamma-3}$ or $\rho\propto
r^{-\gamma}(1+(r/r_s)^{3-\gamma})^{-1}$ with $\gamma=1$, was first
presented by NFW based on a suite of collisionless n-body simulations
with different initial density fluctuation spectra and cosmological
parameters.  More recent work debates the value of $\gamma$
\citep{Moore.etal:98,Jing.Suto:00}, but most estimates are in the
range $1<\gamma<1.5$.  Even before the discovery of a universal
density profile, several authors pointed out the apparent discrepancy
between the cuspy central density profiles predicted by
dissipationless numerical simulations and those inferred from the
rotation curves of galaxies
\citep{moore:94,flores.primack:94,burkert:95}. Although some recent
observational evidence claims a marginal consistency with such cuspy
dark matter profiles \citep{vandenbosch.swaters:01}, most do not
\citep{cote.etal:00,deBlok.etal:01,Blais-Ouellette.etal:01}.

This led to a flurry of papers trying to explain the discrepancy.
Given the numerous other successes of the CDM model most of these
explanations involved altering the model only at small scales, which
would only affect the cuspiness of the dark matter density profile.
Some changed the fundamental nature of the dark matter particle
itself: collisional dark matter \citep{spergel.steinhardt:00},
decaying dark matter \citep{cen:00}, fluid dark matter
\citep{peebles:00}, repulsive dark matter \citep{goodman:00}, and
annihilating dark matter \citep{kaplinghat.etal:00}.  Others changed
the temperature of the dark particle from cold to warm to reduce the
small scale power, i.e.  warm dark matter
\citep{hogan.dalcanton:00,colin.etal:00,eke.etal:01,
  avila-reese.etal:01,bode.etal:01}, while others altered the shape of
the spectrum at small scales by changing inflation
\citep{kamionkowski.liddle:00}. At the most extreme, some authors
suggested that the only solution was to change the nature of gravity
itself \citep{deblok.mcgaugh:98}.

The simulations that predicted the existence of the central dark
matter cusp \citep{Navarro.Frenk.White:97,Moore.etal:98,Jing.Suto:00}
only included the dark matter component.  By using rotation curve
analyses and light profiles, the same techniques used to infer the
above mentioned discrepancy, we know that the central regions of most
galaxies are dominated by a baryonic component
\citep{vanalbada.sancisi:86}.  The dynamics of the baryonic component
can be quite different than that of dark matter owing to its
dissipative nature: baryons can shock and cool.  Furthermore, even a
dissipationless, stellar baryonic component can have a markedly
different geometry than pure dark matter since it originally formed
from a dissipative component.  These different dynamical properties
and the subsequent interactions of the baryonic component with the
dark matter could greatly affect the dark matter structure.

Since it is commonly assumed that the addition of a baryonic component
would cause the dark matter to adiabatically contract
\citep{blumenthal.etal:86} thereby exacerbating the observed
discrepancy, the complications caused by the baryonic component are
usually dismissed.  Even so, both \citet{binney.etal:01} and
\citet{el-zant.etal:01} discuss mechanisms by which the baryonic
component might produce a core.  Much of the baryonic component will
cool and likely form a disk \citep{fall.efstathiou:80}.  Disk bars and
strong grand-design structure can strongly interact with the dark
matter \citep{Weinberg:85,Hernquist.Weinberg:92} by transferring
angular momentum to the spheroid-halo component.  Quoting from
Hernquist and Weinberg (1992)

\noindent
``Cosmological simulations of [dark] halo formation from plausible
initial conditions generally yield halos whose density structure
resembles that of [a Hernquist profile] \citep{dubinski.carlberg:91}.
While somewhat uncertain, it would appear that this result is in
conflict with observed rotation curves of at least some galaxies where
the dark matter halos apparently have significant core regions with
roughly constant density (e.g.  \citet{flores.etal:93}).  The
simulations reported here [of rotating bars] provide, in principle, a
mechanism for developing cores in halos long after they form.''

\noindent
It is this mechanism that we explore in this paper.  We present a
brief description and simulations of the mechanism in \S\ref{sec:mech}
and in \S\ref{sec:disc} we discuss the dynamics in more detail along
with other implications.

\section{The Mechanism}
\label{sec:mech}

In an equilibrium dark matter halo the average orbit is highly
eccentric ($e\approx0.5$ on average) and, therefore, has low angular
momentum for its energy.  Very little energy is required to remove the
cusp; a minor addition of angular momentum provides enough
circularization to exclude an eccentric orbit from the central region.
The pattern of a rotating bar has enough angular momentum, if
transferred to the cusp, to remove it altogether.  Our basic picture
is as follows:
\begin{enumerate}

\item Start with a forming galaxy.  CDM simulations predict that its
  dark matter distribution will be cuspy (e.g. an NFW profile).
  
\item As the galaxy continues to form, smooth accretion of gas and
  stochastic merging and dissipation is followed by the settling of the
  baryons into a mostly gaseous, cold disk.
  
\item As the disk becomes more massive and centrally concentrated, at
  some point it will overwhelm the halo support and form a bar as seen
  in simulations of galaxy formation that include a dissipative
  baryonic component \citep{katz.gunn:91,steinmetz:97}.  Since the
  disk is mostly gaseous and cold and the halo is quite disturbed, it
  makes it easy to form a very strong bar.  Simulations suggest that
  for Milky Way sized galaxies
  early bars have semi-major axes approaching 10 kpc \citep{steinmetz:97}.
  
\item The rotating bar pattern excites a gravitational wake in the
  dark matter.  The wake trails the bar causing the bar to slow its
  rotation and transfer its angular momentum to the dark matter
  \citep{Weinberg:85}.  This forms a core in the dark matter
  distribution.  The 10 kpc primordial bar will remove the cusp out
  to $\sim$5 kpc in $\sim$1.5 gigayears.
  
\item After the formation of the core early on, a {\em classic}
  stellar bar may form and experience low torque in the current epoch.

\end{enumerate}

It is easy to estimate that a strong bar can have important
consequences for halo evolution.  A non-axisymmetric bar force is
dominated by its quadrupole component.  A toy model for a rotating
gravitational quadru\-pole is two masses in orbit at the same distance
from the center of a galaxy but at opposite position angles.  We can
use the Chandrasekhar dynamical friction formula to estimate the time
scale for transferring all of the bar's angular momentum to the halo.
The answer is a few bar rotation periods \citep[see][]{Weinberg:85}.
Since the total angular momentum in the bar approaches that of the
dark matter halo within the optical radius, we conclude that the
evolution of the dark matter halo induced by a bar can be significant.
These estimates have been further refined by
\citet{Hernquist.Weinberg:92} and
\citet{Debattista.Sellwood:98,Debattista.Sellwood:00} using n-body
simulations.

The evolution of the halo may be estimated analytically using a
perturbation expansion of the collisionless Boltzmann equation and a
solution of the resulting initial value problem.  The zeroth-order
solution specifies the equilibrium galaxy and the first-order solution
determines the linear response of the galaxy to some perturbation.
The second-order solution contains the first non-transient change in
the underlying distribution.  By taking the limit for the evolution
time scale to be much larger than an orbital time, the transient
contribution can be made arbitrarily small.  However, we are often
interested in intermediate time scales.  Explicit comparisons with
n-body simulations shows that this approximation is acceptable even
for a small number of orbital time scales.  Mathematical details can
be found in \citet{Weinberg:85,Weinberg:01a}.

\begin{figure*}
  \subfigure[Linear theory]{\includegraphics[width=3.5in]{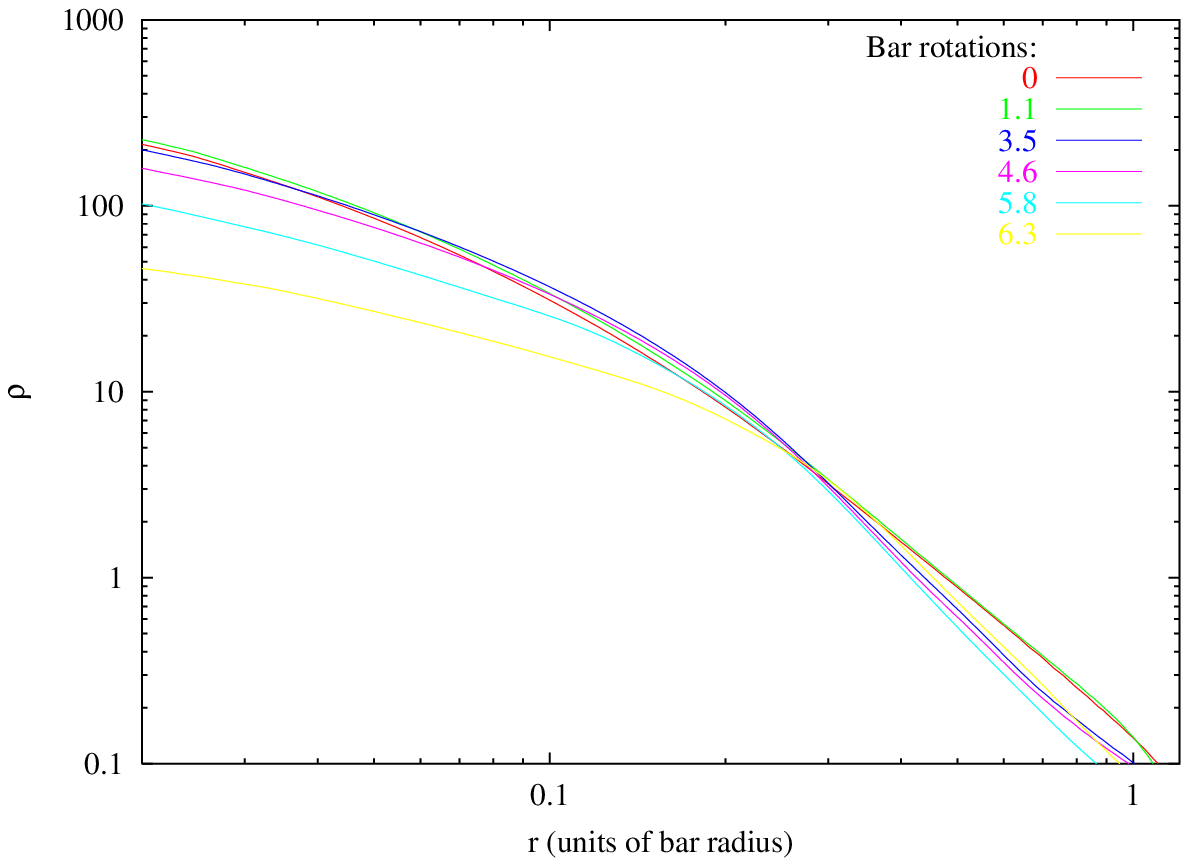}}
  \subfigure[Simulation]{\includegraphics[width=3.5in]{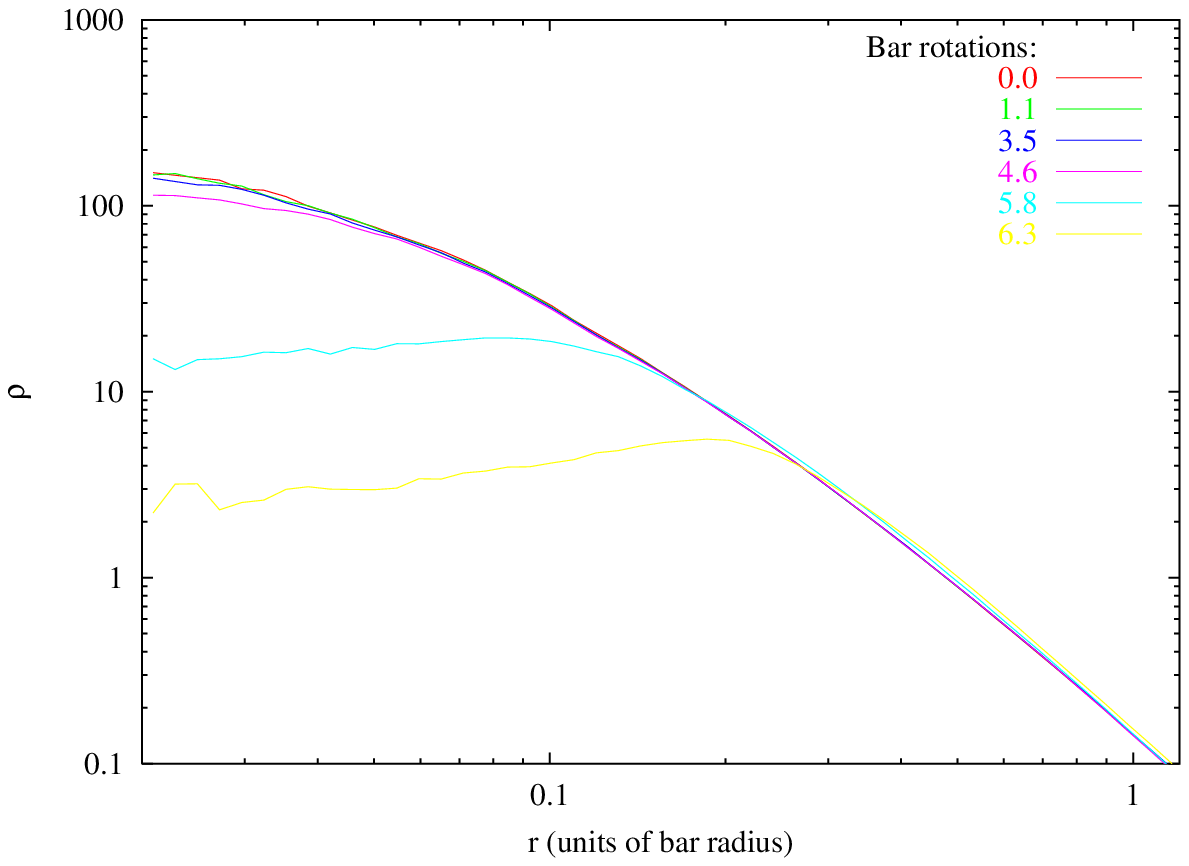}}
  \caption{
    Evolution of an NFW profile with an embedded disk bar.  Profiles
    are a time sequence labeled by the number of bar rotation times as
    predicted from linear perturbation theory (a) and from a n-body
    simulation (b).}
  \label{fig:barNFW} 
\end{figure*}

Figure \ref{fig:barNFW}a shows the perturbation theory prediction for
the evolution of the density profile.  The simulation is of a bar in
an NFW profile with a concentration of 20.  The disk contains half the
mass within the bar radius and the bar contains 30\% of the disk mass.
We choose the corotation radius to be the NFW scale length and the bar
radius is chosen to be 0.5 scale lengths.  The results described below
are only weakly sensitive to this choice.  The bar figure is
represented by a homogeneous ellipsoid with axis ratios of 1:0.5:0.05.
We derive the bar force from the quadrupole term of the gravitational
potential of the ellipsoid.  Ignoring the higher order multipoles in
the bar potential does not greatly change the results but does cause
us to slightly underestimate the evolution
\citep{Hernquist.Weinberg:92}.

The angular momentum transfer to the dark matter takes place at
commensurabilities between the pattern speed and orbital frequencies
and is dominated by a few strong resonances.  The torque pushes the
inner halo orbits to higher angular momentum and energy, removing
gravitational support for the inner cusp. The halo expands and
flattens the cusp.  After a few bar rotation times (several hundred
million years) the central cusp is flattened, but not enough to match
the observations.

Although the perturbative approach yields more accurate results, free
from the numerical noise inherent in n-body simulations, they are only
valid in the linear regime.  Hence, we perform n-body simulations
using a parallel implementation with the Message Passing Interface
(MPI), of the algorithm described in \citet{Weinberg:99}, a self
consistent field (SCF) technique.  This algorithm defines a set of
orthogonal functions whose lowest-order member is the unperturbed
profile itself.  Each additional member in the series probes
successively finer scales.  Because all scales of interest here can be
represented with a small number of degrees of freedom, the particle
noise is low.

The initial conditions are a Monte Carlo realization of the exact
isotropic phase space distribution function for the NFW profile,
determined by Eddington inversion \citep[see][Chap.
4]{Binney.Tremaine:87}, using 4,000,000 equal mass particles.  As we
discuss below this number of particles is sufficient to give a
converged result.  Furthermore, when simulated without a bar
disturbance this realization does not evolve and remains in
equilibrium.  The rotating bar disturbance, with the same strength and
size as in the perturbative calculation, is again represented by the
quadrupole term of gravitational potential of the ellipsoid.  It is
turned on adiabatically over four bar rotation times to avoid sudden
transients.  The use of the quadrupole term only, the first
contributing multipole after the monopole, ensures that the dark
matter halo remains in approximate equilibrium as the bar perturbation
is applied.  The early n-body evolution, shown in Figure
\ref{fig:barNFW}b, is similar to the results from the perturbative
approach.  The inner evolution follows the analytic predictions up to
five bar rotation times.  At this point, approximately 30\% of the
available angular momentum in the bar pattern has been transferred to
the halo.  Subsequent evolution is more rapid, presumably due to the
non-linear response of the near-resonant orbits, although the details
remain to be investigated. A similar super-linear increase in torque
was reported by \citet{Hernquist.Weinberg:92}.  The cusp within $\sim
0.5$ bar radii is completely removed after only 7 bar rotations.  At
this time the bar pattern has lost all of its original angular
momentum, the stars in the bar have lost $\sim 25$\% of their angular
momentum, and the disk as a whole has lost only $\sim$8\% of its
original angular momentum, and the disk as a whole has lost only
$\sim$8\% of its original angular momentum.

\section{Discussion}
\label{sec:disc}

\subsection{Dynamics}

In the previous section we mentioned that one can estimate the
dynamical friction decay time of a rotating bar by applying the
Chandrasekhar dynamical friction formula to a toy model for the bar
consisting of two masses in orbit at the same distance from the center
of the galaxy but at opposite position angles.  This physical argument
is a gross simplification, however.  The Chandrasekhar dynamical
friction formula is derived by considering the momentum transfer of
stars gravitationally scattered by a traveling body.  The problem is
much more complicated in a halo where individual orbits are
quasiperiodic.  The gravitational wake, i.e. the response of the dark
halo to the bar, is not as simple as the Chandrasekhar formula would
lead one to believe.  A given orbit may encounter the rotating
perturbation many times.  If the precession frequency leads or trails
the pattern, the net torque applied to this orbit will be zero on
average.  This appears to be in conflict with dynamical friction but
really it is not.  The bar will only receive a net torque if the
gravitational wake either trails or leads the bar pattern and this can
only occur if some of the orbital actions are changed, something that
can only occur at or near resonances.  At exact commensurabilities
between the orbital and pattern frequencies individual orbits receive
kicks breaking symmetry, resulting in a density response that trails
the bar.  To have this symmetry broken requires gradients in the phase
space density at the positions of these resonances.

\begin{figure*}
  \epsscale{1.2}
  \plotone{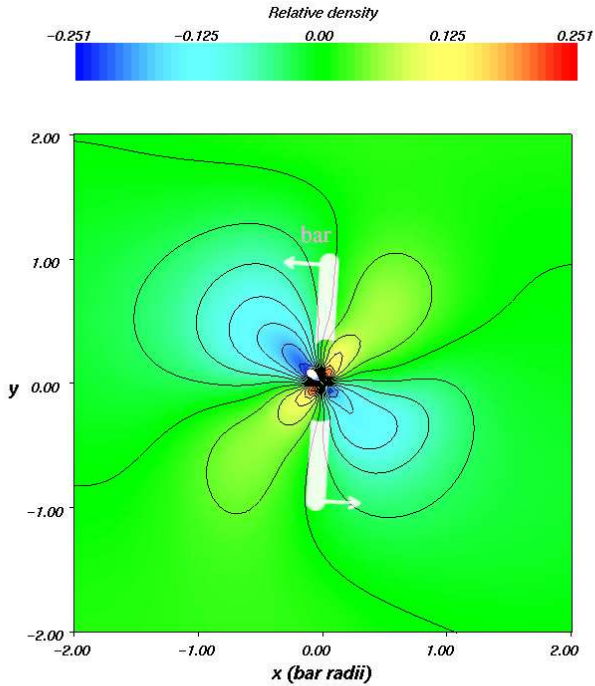}
  \caption{The halo response and bar position during the evolution
    of an NFW model halo.  We subtract the mean density and plot the
    amplitude of the resulting wake where white  represents underdense regions
    and black overdense regions.}
  \label{fig:halo_wake}
\end{figure*}

In Figure \ref{fig:halo_wake}, we plot the actual gravitational wake
induced by a bar in an n-body simulation.  It is similar to the
simulations described above but uses an NFW halo realized with $10^7$
equal mass particles and is plotted after 5 bar pattern rotations.
This figure shows the halo density distortion in response to the bar.
The bar's size and phase are shown schematically.  The bar pattern
leads the wake and therefore the bar torques the halo.

The location in the halo of the peak angular momentum transfer depends
on the halo profile itself in two ways.  First, the torqued orbits
will be near commensurabilities between the orbital frequencies and
the pattern speed.  These commensurabilities or {\em resonances} take
the form $l_r\Omega_r + l_\phi\Omega_\phi = m\Omega_{bar}$ where the
three values of $\Omega$ are the radial, azimuthal and pattern
frequencies, respectively, $l_r$ and $l_\phi$ are integers and $m$ is
the azimuthal multipole index.  For the $l=m=2$ multipole, we will
denote a particular resonance by the tuple $(l_r, l_\phi)$.  Low-order
(high-order) resonances have small (large) values of $|l_r|$ or
$|l_\phi|$.  Second, a net torque requires a differential in phase
space density on either side of a particular resonance.  It is not the
dark matter density of the inner halo but a large phase space gradient
that is key to a large bar torque.  In fact, if the bar were located
in an infinitely large homogeneous core, no matter how dense, there
would be no net torque, in contrast to the predictions of the simple
toy model.  If the bar is inside the halo core, then there will be
very few nearby resonances with a differential in phase space density
and the torque would be diminished; the dominant resonant orbits in
this case would be at or beyond the core radius.  Conversely, if the
bar is in a dark matter cusp, orbits near and inside the bar radius
cover a large range of frequencies. There are low-order resonances
deep in the cusp where the phase space gradient is large.  The torqued
cusp orbits move to larger radii, decreasing the cusp density and
decreasing the overall depth of the potential well.  Both of these
effects cause the cusp to expand overall.  Thus, the formation of a
bar will naturally eliminate an inner cusp.

\begin{figure*}
  \epsscale{1.0}
  \plotone{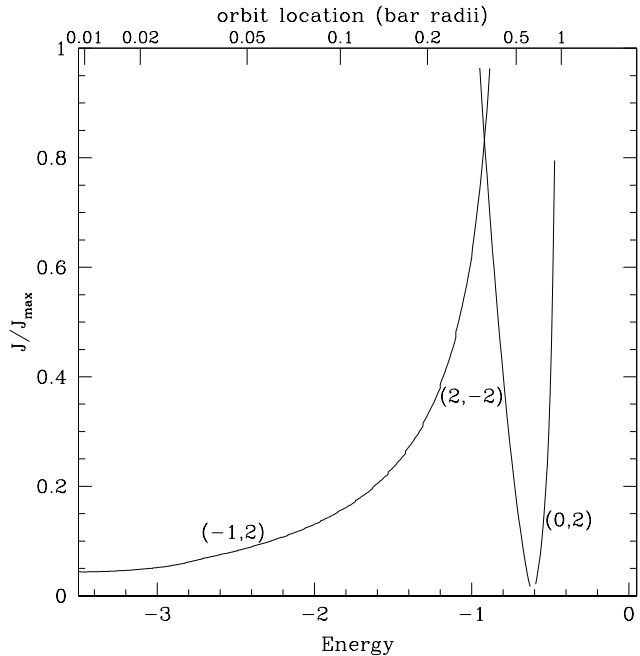}
  \caption{Location of low-order resonances in energy (lower axis) and
    characteristic radius (upper axis) for the bar in the NFW profile.
    The vertical axis describes the orbital angular momentum $J$ in
    units of the maximum angular momentum for a given energy,
    $J_{max}$.  The inner Lindblad-like resonance extends throughout
    the inner cusp.  This resonance is absent from dark halo
    models with cores.}
\label{fig:resloc}
\end{figure*}

Linear theory allows us to explicitly identify the dominant resonances
and the angular momentum transfer to halo orbits for models both with
and without cores.  The overall torque is dominated by one or two
resonances in each case.  For the NFW profile, the torque is dominated
by the resonance $(-1, 2)$; this is analogous to an inner Lindblad
resonance.  This resonance does not occur for a comparable
\citet{King:66} model.  The first contributing resonance for the King
model is $(2,-2)$ near the location of the core radius.  Higher-order
resonances that occur at larger radii are well confined to a single
characteristic radius (close to vertical in Fig. \ref{fig:resloc}.
However, the $(-1,2)$ resonance is unique.  It always exists in a cusp
as $r\rightarrow0$ because $\Omega_r\rightarrow 2\Omega_\phi$ as the
orbital angular momentum $J$ approaches zero. Therefore there is
always some value of $J$ for which $-\Omega_r + 2\Omega_\phi =
2\Omega_{bar}$ even though $\Omega_r$ and $\Omega_\phi$ both diverge
for small $r$.  For this reason, the $(-1, 2)$ resonance can affect
orbits deep within a cusp, dramatically changing the inner profile.
For a model with a core, the core expands somewhat as angular momentum
is transferred toward the outer halo but otherwise remains
qualitatively similar to its initial state.

Since bar instability is ubiquitous in self-gravitating disks,
continued accretion and cooling is likely to precipitate a bar
instability in the forming inner gas disk early on.  Once the rotating
bar forms, its body angular momentum can be transferred to the halo as
described above, flattening and removing the inner cusp.  It is
possible for this process to occur in stages with multiple bars; at
each stage the inner core will grow.  There will be a transition in
the magnitude of the torque and a slowing of halo evolution when the
inner $(-1,2)$ resonance is finally eliminated.

\subsection{Implications}

Because the bar--cusp coupling depends on near-resonant dynamics,
simulations with very high resolution will be required to resolve the
dynamics properly.  We are confident in our n-body results since they
agree with the exact perturbative results in the linear regime, once
we have enough particles to reach convergence.  To test for
convergence we ran a suite of simulations with increasing particle
number from $10^4$ through $4 \times 10^7$ and determined that, for
our SCF expansion method, we obtained convergence in the evolution for
particle numbers $\gtrsim4\times10^6$.  This class of potential solver
\citep[see][]{Clutton-Brock:72, Clutton-Brock:73, Kalnajs:76*3,
  Fridman.Polyachenko:84*3, Hernquist.Ostriker:92} suppresses small
scale fluctuations.  Direct summation approaches
\citep[e.g.][GRAPE]{GRAPE:00}, tree codes \citep{Barnes.Hut:86} and
grid based codes \citep{sellwood.merrit:94,pearce.couchman:97} have
more inherent small scale noise and will most likely require even
higher particle numbers to obtain the same convergence.  Note that
even if the resonances are not resolved, these simulations will still
exhibit significant torque.  Our suite of simulations shows that the
overall torque {\em increases} as the particle number {\em
  decreases}\footnote{If the orbit diffusion becomes large enough, it
  is plausible that the torque will decrease to zero, however.}!
These same simulations give us good agreement with Chandrasekhar's
formula.  Such agreement is not a good indication that one is
observing the correct dynamics.  Conversely, the Chandrasekhar formula
works well in simulations with low to moderate particle number because
the resonances are obliterated by artificial diffusion and, therefore,
are well represented by simple scattering.  In short, it is difficult
to see resonant effects in n-body simulations because the diffusion
rate is high for moderate numbers of particles.  Astronomical sources
of noise such as orbiting substructure, decaying spiral waves,
lopsidedness, etc. do not produce enough small scale noise to affect
this resonant evolution but instead produce large scale deviations
from equilibrium that will not drive significant relaxation in the
inner halo within several orbital times \citep[see][for estimates of
these timescales]{Weinberg:01a, Weinberg:01b}.

The evolution of bars within dark halos has recently been studied by
\citet{Debattista.Sellwood:98, Debattista.Sellwood:00}.  They
simulated bars in dark halos by constructing a strong bar from a
Q-unstable disk and then following its evolution.  They found that the
bars rapidly transferred angular momentum to the massive
(non-rotating) halos as predicted by \citet{Weinberg:85}.  On the
other hand, a few direct and a wider variety of indirect inferences
from observations indicate that galaxies today harbor rapidly rotating
bars.  Hence, Debattista \& Sellwood concluded that dark matter halos
must have low density (or large cores) to be consistent with
observations and could not have cusps.  These results are consistent
with our scenario.  As the disk matures and becomes stellar rather
than gas dominated, a normal stellar bar may form through secular
growth or instability.  The first generation of bar evolution would
have eliminated the inner halo torque by removing the cusp and, hence,
be consistent with the \citet{Debattista.Sellwood:00} arguments.

The bar torque will affect the kinematics of the bulge.
\citet{Kormendy:82} finds that bulges of SB galaxies have sufficient
spin to be rotationally flattened.  However, triaxial galaxy bulges
appear to be kinematically midway between an isothermal and
rotationally-supported disk component.  The bulge length scale is
smaller the {\em classic} bar length and much smaller than the length
of a strong primordial bar.  Because the bulge profile is not
supported by relatively low-energy eccentric orbits, the bulge will
not be subject to the strong density evolution predicted here.

Most of the evidence against central dark matter cusps in galaxies
concerns dwarf galaxies, particularly those with low surface
brightness \citep{cote.etal:00,deBlok.etal:01,Blais-Ouellette.etal:01}
since these systems offer more precise rotation curve decompositions.
One might not expect strong bars to form in such systems and, hence,
for the mechanism proposed here not to have much relevance.  However,
the same analysis used to indicate the lack of a central dark matter
cusp also shows that these low surface brightness galaxies are
deficient in baryons by three or four times compared to normal
galaxies \citep{vandenbosch.swaters:01}.  If a strong bar forms in a
gas rich disk of a dwarf galaxy, much of the gas will lose substantial
amounts of angular momentum and be driven towards the center
\citep{roberts.etal:79}.  This dense cold gas would then be expected
to undergo a strong starburst and, owing to the shallow potential well
of the dwarf galaxy, much of the gas would be expelled as a wind
\citep{dekel.silk:86}.  The remaining galaxy would be one of low
surface brightness possessing a core in its dark matter distribution.
Our bar mechanism, therefore, not only provides a natural explanation
for the existence of dark matter cores but for the existence of low
surface brightness dwarfs as well; those dwarf galaxies that had the
strongest bars will have the largest cores and lowest surface
brightness.  There is also some evidence that larger, high surface
brightness galaxies also have cores in their central dark matter
distributions, although such determinations are much more difficult
\citep{Debattista.Sellwood:00, binney.evans:01}.

Even if our proposed bar mechanism does not explain the lack of
observed central dark matter cusps in all cases, it will still have a
profound effect on the structure and evolution of almost all galaxies.
Hence, both to understand galaxy formation and evolution and to make
predictions from theory it is necessary to resolve these dynamical
processes.  For example: it changes the predicted dark matter
densities at the solar radius and would change the predicted dark
matter detection rates \citep[see][and references
therein]{stiff.etal:01}, it could change the rotation velocities used
in Tully-Fisher predictions and perhaps improve the agreement of
theory with observations \citep{navarro.steinmetz:00}; it could change
disk scale lengths \citep{steinmetz.navarro:99}; and it could make
satellite galaxies easier to tidally disrupt by decreasing their
central densities \citep{moore.etal2:99}.  Some of these consequences
are qualitatively discussed in \citet{binney.etal:01} but it is hard
to address many of these issues quantitatively with the simplified
simulations presented here.  In the future, we plan to perform
simulations with self consistent disks and bars, eventually including
gas dynamics, to investigate these issues.

The lack of an observed central dark matter cusp in galaxies is a
consequence of simple dynamical evolution and does not require a
fundamental change to the nature of dark matter or galaxy formation.
The difference in evolutionary end states may be the result of strong
star formation feedback evacuating the more weakly bound central
potentials, lack of strong accretion events and mergers after the
primordial bar has disappeared, or a combination of the two.  Indeed,
such stochastic processes naturally predict the inferred dispersion in
present day profiles.  However, to correctly resolve these important
dynamical processes in {\it ab initio} calculations of galaxy
formation remains a daunting task, requiring at least 4,000,000 halo
particles using our SCF code, and requiring many times more particles
when using noisier tree, direct summation, or grid based
techniques---the usual methods employed in such calculations.

\acknowledgements We wish to thank Eric Linder, Ariyeh Maller and
David Weinberg for useful discussions.  This work was supported in
part by NSF AST-9802568 and AST-9988146, and by NASA LTSA NAG5-3525.

\end{document}